\documentclass[aps,prl,twocolumn,superscriptaddress]{revtex4-1}
\usepackage[colorlinks=true, citecolor=magenta, linkcolor=blue, urlcolor=blue]{hyperref}
\usepackage{graphicx}
\usepackage{amsmath}
\usepackage{amssymb}
\usepackage{ulem}
\usepackage{amsfonts}
\usepackage{xcolor}
\usepackage{tikz}
\usepackage{pdfpages}
\usepackage{empheq}

\DeclareMathOperator{\im}{Im}
\DeclareMathOperator{\sgn}{sgn}

\allowdisplaybreaks

\makeatletter
\AtBeginDocument{\let\LS@rot\@undefined}
\makeatother

\begin{document}
 
\title{Microscopic Theory of Ultrafast Skyrmion Excitation by Light}
\author{Emil Vi\~nas Bostr\"om}
\affiliation{Max Planck Institute for the Structure and Dynamics of Matter, Luruper Chaussee 149, 22761 Hamburg, Germany}
\author{Angel Rubio}
\affiliation{Max Planck Institute for the Structure and Dynamics of Matter, Luruper Chaussee 149, 22761 Hamburg, Germany}
\affiliation{Center for Computational Quantum Physics (CCQ), The Flatiron Institute, 162 Fifth Avenue, New York, NY 10010, United States of America}
\author{Claudio Verdozzi}
\affiliation{Division of Mathematical Physics and ETSF, Lund University, PO Box 118, 221 00 Lund, Sweden}
\date{\today}

\begin{abstract} 
We propose a microscopic mechanism for ultrafast skyrmion photo-excitation via a two-orbital electronic model. In the strong correlation limit the $d$-electrons are described by an effective spin Hamiltonian, coupled to itinerant $s$-electrons via $s-d$ exchange. Laser-exciting the system by a direct coupling to the electric charge leads to skyrmion nucleation on a 100 fs timescale. The coupling between photo-induced electronic currents and magnetic moments, mediated via Rashba spin-orbit interactions, is identified as the microscopic mechanism behind the ultrafast optical skyrmion excitation.
\end{abstract}

\maketitle


Topological magnetic excitations are of large interest both from a fundamental point of view and for the construction of compact and energy efficient memory and computational devices~\cite{Muhlbauer09,Yu10,Yang16,Chisnell15,Cheng16,Nakata17a,Nakata17b,Kasahara18}. A notable example is magnetic skyrmions, topological spin configurations stabilized through a competition of exchange, Dzyaloshinskii-Moriya (DM) and Zeeman interactions, that have generated much attention due to their potential for realizing race track memories~\cite{Sampaio13,Nagaosa13,Romming15} and quantum computation devices~\cite{Yang16,Chauwin19,Rex19}. To exploit skyrmions for technological purposes requires efficient ways of writing, deleting, and manipulating such excitations on short time scales and with high spatial precision. As demonstrated in recent experiments, a promising method to create small skyrmion clusters is by irradiating a chiral magnetic film with femtosecond light pulses~\cite{Finazzi13,Ogawa15,Je18}. However, while several works have addressed laser-induced skyrmion excitation~\cite{Fujita17,Berruto18,Khoshlahni19,Polyakov20}, an adequate understanding of the inherent microscopic mechanism is still lacking. 

One proposed explanation relies on local laser-induced heating described via a stochastic Landau-Lifshitz equation~\cite{Brown63}; this leads to thermal skyrmion excitation within  $0.1 - 1$ ns~\cite{Fujita17,Berruto18,Je18} but misses the microscopic features underlying spin lattice heating. A second mechanism invokes the inverse Faraday effect (IFE), where the laser electric field couples to the system's magnetization via the asymmetric imaginary part of the dielectric tensor~\cite{Ogawa15,Khoshlahni19}. However, the IFE is inefficient for thin magnetic films because of the short propagation length and small associated Faraday rotation. A third proposal considers electromagnetic vortex beams carrying orbital angular momentum and a net magnetic field~\cite{Polyakov20}. However, this mechanism does not explain skyrmion excitation by linearly and circularly polarized laser pulses, as typically employed in experiments.

All above approaches address skyrmion photo-excitation via spin-only descriptions. However, another key
element is the interaction with itinerant electrons. This interaction is ubiquitous in real materials, may lead to a considerably enhanced light-matter coupling compared to direct magneto-optical effects, and can have a significant influence on the dynamics of magnetic systems~\cite{Mentink15,Stepanov18,Claassen17,Bostrom19,Bostrom20}. Furthermore, an explicit account of the role of the conduction electrons in the microscopic mechanism of optical skyrmion generation is expected to be crucial to fully
exploit material-tailoring techniques to optimize the skyrmion excitation probability (using e.g. optical driving or Moir\'e twisting~\cite{McIver19,Sato19,Kerelsky19,Wang20}).

In this work, we address dynamical electronic effects on skyrmion nucleation and propose a microscopic mechanism for ultrafast skyrmion excitation via laser excitation of the electrons of the magnetic material. We consider a two-band model of itinerant $s$- and strongly correlated $d$-electrons in two dimensions, and describe the magnetic moments of the $d$-electrons in terms of an effective spin Hamiltonian. We find that $(i)$ when the $s-d$ exchange is the largest energy scale, skyrmions are excited on a $100$ fs timescale. $(ii)$ The equilibrium phase diagram shows competing spin spiral, ferromagnetic (FM), and skyrmion crystal regions, with a small energy barrier to excite skyrmions from the FM state. $(iii)$ The spin-electron coupling mediated via spin-orbit interactions among the itinerant electrons is essential for skyrmion photo-excitation. Our model provides a proof-of-principle for electronically mediated skyrmion photo-excitation, and is straightforward to realize with standard laser sources.


{\it Model.--}
We consider a system of coupled $s$- and $d$-electrons on a two-dimensional square lattice subject to laser irradiation. The light-matter coupling is described via Peierls substitution where the Peierls phases are $\theta_{ij}(t) = -(e/\hbar) \int_{{\bf r}_j}^{{\bf r}_i} d{\bf r}\cdot {\bf A}({\bf r},t)$ and ${\bf A}({\bf r},t)$ is the vector potential. The total Hamiltonian is $H(t) = H_s(t) + H_d(t) + H_{s-d}$,
where
\begin{align}
 H_s(t) &= \sum_{i\sigma} \epsilon_{i\sigma}(t) \hat{n}^s_{i\sigma} - {\bf B} \cdot \sum_i \hat{\bf s}_i, \label{Ham_s} \\
 &+ \sum_{\langle ij\rangle \sigma\sigma'} e^{i\theta_{ij}(t)} c_{i\sigma}^\dagger (-t{\bf 1} + \boldsymbol\alpha_{s,ij} \cdot \boldsymbol\tau)_{\sigma\sigma'} c_{j\sigma'} \nonumber \\
 H_d(t) &= U_0 \sum_i \hat{n}^d_{i\uparrow}\hat{n}^d_{i\downarrow} + \sum_{\langle ij\rangle} \Big( \frac{V_0}{2} \hat{n}^d_i \hat{n}^d_j - J_0 \hat{\bf S}_i \cdot \hat{\bf S}_j \Big) \label{Ham_d} \\
 &\hspace*{-0.8cm}+\sum_{\langle ij\rangle\sigma\sigma'} e^{i\theta_{ij}(t)} d_{i\sigma}^\dagger (-t_d{\bf 1} + \boldsymbol\alpha_{d,ij} \cdot \boldsymbol \tau)_{\sigma\sigma'} d_{j\sigma'} - {\bf B} \cdot \sum_i \hat{\bf S}_i, \nonumber \\
 H_{s-d} &= t_{s-d} \sum_{i\sigma} \left(c_{i\sigma}^\dagger d_{i\sigma} + d_{i\sigma}^\dagger c_{i\sigma}\right) \label{Ham_sd} \\
 &+ U_{s-d} \sum_{i\sigma\sigma'} \hat{n}^s_{i\sigma} \hat{n}^d_{i\sigma'} - J_{s-d} \sum_{i\sigma} \hat{\bf s}_i \cdot \hat{\bf S}_i. \nonumber
\end{align} 

In Eqs.~(\ref{Ham_s}-\ref{Ham_sd}), $c_{i\sigma}^\dagger$ ($d_{i\sigma}^\dagger$) creates an $s$- ($d$-) electron at site $i$ with spin projection $\sigma$, and $\hat{n}^a_{i\sigma}$ is the spin density operator for orbital $a \in \{s,d\}$ at site $i$. The orbital energy is given by $\epsilon^a_{i\sigma}$, $t_a$ is the hopping amplitude between nearest-neighbor sites $i$ and $j$, and $\boldsymbol\alpha_{a,ij}$ accounts for Rashba spin-orbit interactions. The $s$- and $d$-electron spin operators are given by $\hat{\bf s}_i = c_{i\sigma}^\dagger \boldsymbol\tau_{\sigma\sigma'} c_{i\sigma'}$ and $\hat{\bf S}_i = d_{i\sigma}^\dagger \boldsymbol\tau_{\sigma\sigma'} d_{i\sigma'}$, where $\boldsymbol\tau$ denotes the vector of Pauli matrices and repeated spin indexes are summed over.

In $H_d$, both a local interaction $U_0$ as well as nearest-neighbor direct and exchange interactions $V_0$ and $J_0$ are included~\footnote{The Heisenberg term comes from writing the Coulomb exchange operator as $d_{i\sigma}^\dagger d_{i\sigma'} d_{j\sigma'}^\dagger d_{j\sigma} = (n_i n_j)/2 + 2\hat{\bf S}_i \cdot \hat{\bf S}_j$, and absorbing the first term by a renormalization of $V_0$.}. In the $s-d$ interaction term [Eq.~(\ref{Ham_sd})], $t_{s-d}$ is the hybridization strength, and $U_{s-d}$ and $J_{s-d}$ the direct and exchange interactions. For $\langle \hat{n}^d_i \rangle = 1$ (as assumed in the rest of this work), the direct term just renormalizes the orbital energy $\epsilon_{i\sigma}$. Finally, both $s$- and $d$- electron spins interact with an external static magnetic field ${\bf B}$ via Zeeman coupling.


{\it Effective spin Hamiltonian.--}
For a strongly correlated and half-filled $d$-band with $U_0 \gg \max(t_d, t_{s-d})$, empty and doubly occupied $d$-orbitals can be projected out by a time-dependent Schrieffer-Wolff transformation \cite{Schrieffer66,Eckstein17} (see the SM). To second order in $t/U$ we find
\begin{align}\label{eq:spin_hamiltonian}
 H_d + H_{s-d} &= \sum_{\langle ij\rangle} \Big[J_{ij}(t) \hat{\bf S}_i \cdot \hat{\bf S}_j + {\bf D}_{ij}(t) \cdot (\hat{\bf S}_i \times \hat{\bf S}_j)\Big] \nonumber \\
 &- g \sum_{i} \hat{{\bf s}}_i \cdot \hat{{\bf S}}_i - {\bf B} \cdot \sum_i \hat{\bf S}_i,
\end{align}
while $H_s(t)$ remains unchanged. Here, the first and second terms represent Heisenberg exchange and Dzyaloshinskii-Moriya~\cite{Dzyaloshinskii58,Moriya60} (DM) interactions, respectively, and the third term is a renormalized $s-d$ exchange. The effective spin parameters are respectively given by $J_{ij}(t) = 4t_d^2 I_{ij}(t) - J_0$, ${\bf D}_{ij}(t) = 8it_d \boldsymbol\alpha_{d,ij} I_{ij}(t)$, 
and $g = J_{s-d} - 4t_{s-d}^2/U$, where 
\begin{align}\label{eq:integral}
 I_{ij}(t) &= \im \int_{-\infty}^t d\bar{t}\, e^{iU(t-\bar{t})} e^{0^+ \bar{t}} \cos(\theta_{ij}(t)-\theta_{ij}(\bar{t})).
\end{align}
In equilibrium, where $\dot{\theta}_{ij}(t) = 0$, the integral reduces to the well-known expression $1/U$, with $U = U_0 - V_0$. However, out of equilibrium, Eq.~(\ref{eq:spin_hamiltonian}) shows that exchange and DM couplings can be manipulated by varying the frequency and field strength of the laser~\cite{Mentink15,Stepanov18,Claassen17,Bostrom20}. In what follows we assume the dynamic renormalization of the spin parameters is small, and neglect the contribution to the Peierls phases from the external magnetic field~\footnote{A Floquet analysis shows that $I_{ij}(t)$ deviates significantly from its static value only for $n\hbar\omega = U$, with $n$ integer, or for very large electric fields. Also, including the contribution to the Peierls from  external magnetic field {\bf B} would lead to an enlarged magnetic unit cell with corresponding Landau levels for the itinerant electrons. However, for the magnetic fields considered here the ratio $\Phi/\Phi_0$ of the magnetic flux to the flux quantum $\Phi_0 = e/2h$ is very small throughout the system (of the order of $10^{-5}$), and can thus be neglected. In the effective spin Hamiltonian describing the $d$-electrons there is no effect, since any static contribution to the Peierls phases cancels out (see Eq.~\ref{eq:integral})}.


{\it Equations of motion.--}
We now take the semi-classical limit $\hat{\bf S}_i \rightarrow  \langle \hat{\bf S}_i \rangle \equiv {\bf S}_i$, which is exact only for $S \to \infty$ but in practice works well for systems with large spins~\cite{Lieb73,Fradkin13,Heinze11}. Within the semi-classical approximation, the present model corresponds to a two-dimensional generalization of the model introduced in Ref.~\cite{Bostrom19}. Using the Heisenberg equations of motion for the spin operators and defining  ${\bf n}_i = {\bf S}_i/S$ (with $S = |{\bf S}_i|$), the spin dynamics are governed by the Landau-Lifshitz equation
\begin{align}\label{eq:spin_force} 
 \frac{\partial {\bf n}_i}{\partial t} &= {\bf n}_i \times \Big( \sum_{\langle j\rangle} S\big[J_{ij} {\bf n}_j + {\bf D}_{ij} \times {\bf n}_j\big] + {\bf B} + g \langle \hat{\bf s}_i \rangle\Big),
\end{align}
where the last term couples the classical spins to the quantum averages of the $s$-electron spins. The quantum dynamics of the $s$-electrons and their coupling to the classical spins are described via the electronic single-particle density matrix
\begin{align}\label{eq:eom}
 \frac{d}{dt}\rho_{ij}(t) +i\left[H_s(t) + H_{s-d}({\bf n}(t)), \rho(t)\right]_{ij} = 0.
\end{align}
Within the quantum-classical scheme of Eqs.~(\ref{eq:spin_force}-\ref{eq:eom}), connections to electronic reservoirs and interactions among the $s$-electrons can easily be included via non-equilibrium Green's functions~\cite{Petrovic18,Bostrom19}. However, since this increases the computational effort, we here for simplicity take our system as isolated and consider non-interacting $s$-electrons.


{\it Creating a skyrmion with light.--} The dynamics of the system is initiated by an electric field given by ${\bf E}({\bf r},t) = E e^{-r^2/2\sigma^2} \sin^2 \left(\frac{\pi t}{2\tau} \right) \sin(\omega t) \hat{\bf x}$ for $t \in (0, 2\tau)$ and zero otherwise. The pulse width is set by $\tau = 60$ fs, giving a full width at half maximum of $30$ fs, and the photon energy is $\hbar\omega = 0.5$ eV. The distance $r$ is measured from the system center, and the spot size of the laser is controlled by the parameter $\sigma = 6a$ where $a$ is the lattice spacing. Taking $a = 10$ {\AA } and a field strength of $E = 10^9$ V/m, the pulse has an integrated fluence $F = 2.9$ mJ/cm$^2$ similar to typical experiments~\cite{Je18}.

For the spin parameters we take $J = 50$ meV, $D = 25$ meV and $B = 10$ meV, consistent with the values for the prototypical B20 chiral magnet FeGe~\cite{Yin16,Grytsiuk19} at a lattice parameter $a = 10$ \AA. We include a small Gilbert damping $\alpha_G = 0.01$ in the Landau-Lifshitz equation to mimic a dissipative environment and to stabilize the dynamics at long times. The parameters for the $s$-electrons are chosen as $t = 1$ eV, $\alpha = 0.5$ eV and $\epsilon_{i\sigma} = -B\sgn{\sigma}$. Finally, since typically the direct ferromagnetic exchange is larger than the $s-d$ hybridization induced antiferromagnetic exchange~\cite{Zener51,Anderson55}, we take $g = J_{s-d} - 4t_{s-d}^2/U > 0$. 

\begin{figure}
 \includegraphics[width=\columnwidth]{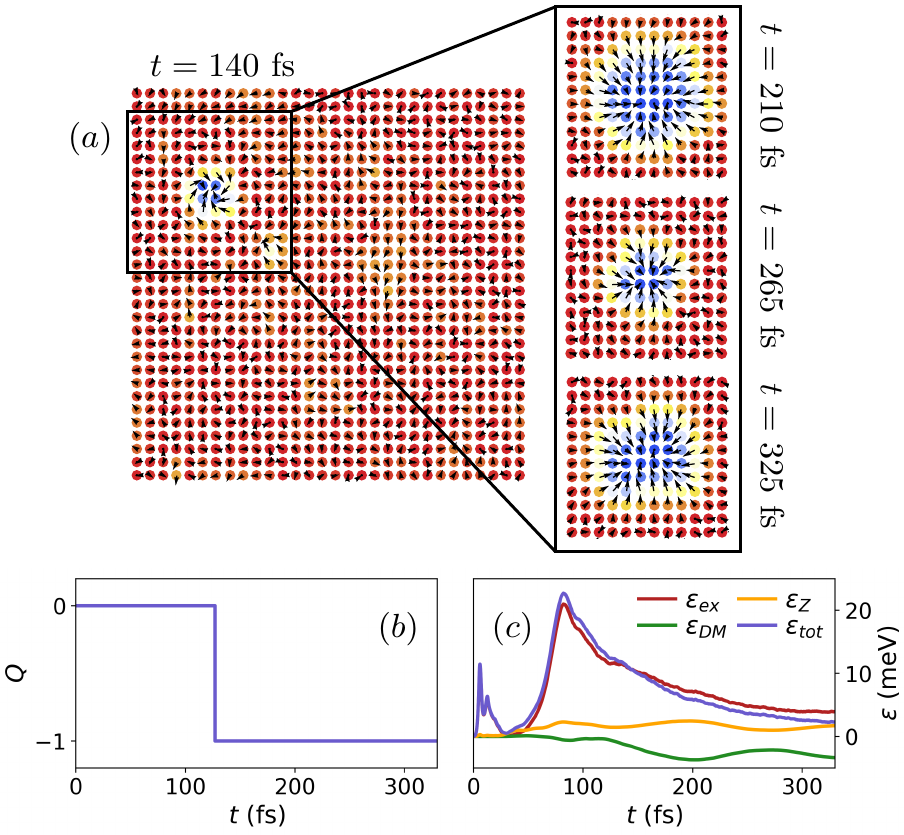}
 \caption{{\bf Laser-induced skyrmion excitation}. $(a)$ Spin configurations at $t = 140$ fs (just after skyrmion excitation), $210$ fs, $265$ fs and $325$ fs. The coloring shows the out-of-plane component of the spin vector ranging from $S_z = -1$ (blue) to $S_z = 1$ (red). The system has $30 \times 30$ sites and the spin parameters are $J = 50$ meV, $D = 25$ meV and $B = 10$ meV. The electronic parameters are $t = 1$ eV and $\alpha = 0.5$ eV, and the spin-electron coupling $g = 2.5$ eV. The laser has a field strength $E = 10^9$ V/m, frequency $\hbar\omega = 0.5$ eV, pulse length $\tau = 30$ fs, and spot size $\approx 12$ nm. $(b,c)$ Topological charge $Q$ and the energy per spin $\epsilon = E/N$ as a function of time. The total energy $\epsilon_{tot}$ is decomposed into exchange, DM and Zeeman contributions $\epsilon_{ex}$, $\epsilon_{DM}$ and $\epsilon_Z$.}
 \label{fig:skyrmion_creation_light}
\end{figure}

In Fig.~\ref{fig:skyrmion_creation_light} we show the photo-induced excitation of a N\'eel skyrmion in a system with $30 \times 30$ sites (for Rashba spin-orbit the induced DM coupling favors N\'eel skyrmions). During the subsequent evolution the skyrmion remains stable but its radius oscillates, indicating that the skyrmion is created in a breathing mode~\cite{Onose12,Mochizuki12}. The topology of the spin texture is quantified via the lattice topological charge $Q = \sum_i Q_i$ (see SM), where $Q_i = \mp 1$ for a single skyrmion (anti-skyrmion). As seen in Fig.~\ref{fig:skyrmion_creation_light}$(b)$, laser irradiation leads to the excitation of a single skyrmion after a time $t \approx 127$ fs, where the topological charge suddenly becomes $Q = -1$. 

\begin{figure}
 \includegraphics[width=\columnwidth]{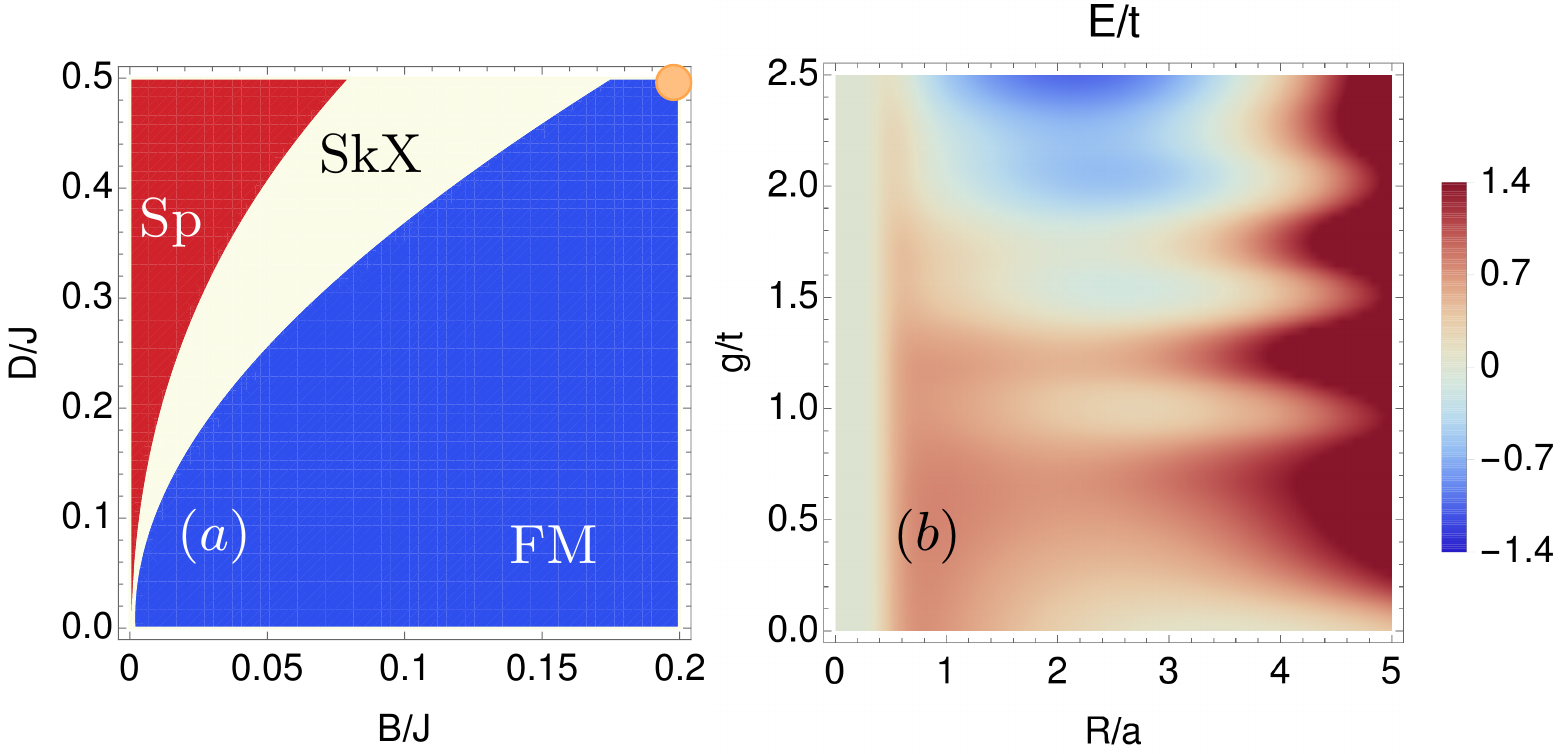}
 \caption{{\bf Equilibrium magnetic properties}. $(a)$ Magnetic phase diagram of an isolated spin system ($g = 0$) as a function of external magnetic field $B$ and Dzyaloshinskii-Moriya interaction $D$. Blue areas indicate a ferromagnetic (FM) state, red areas a spiral (Sp) state and white areas a skyrmion crystal (SkX) state. $(b)$ Excitation energy $E = E_{Sk} - E_{FM}$ as a function of skyrmion radius $R$ and spin-electron coupling $g$, for an exchange interaction $J = 50$ meV, DM interaction $D/J = 0.5$, magnetic field $B/J = 0.2$, hopping $t = 1$ eV and spin-orbit interaction $\alpha = 0.5$ eV. The orange dot in $(a)$ indicates the values of the spin parameters used in $(b)$.}
 \label{fig:phase_diagram}
\end{figure}

As discussed in Refs.~\cite{Je20,CortesOrtuno17,Malottki19} and illustrated in Fig.~\ref{fig:phase_diagram}$(b)$, skyrmions in discrete systems experience no real topological stability, and are in practice only protected by an energy barrier. By looking at the energy per spin $\epsilon = E/N$ [Fig.~\ref{fig:skyrmion_creation_light}$(c)$], we note that it takes the system about $100$ fs to transfer energy from the itinerant electrons to the spins and overcome the barrier. Therefore, the excitation of the spin system is delayed with respect to the electronic system, and the skyrmion is created some time after the pulse has passed.

The demonstration of photo-induced skyrmion creation on ultrafast timescales by excitation of the electronic subsystem constitutes the central result of this work. We note that that to pin down the factors contributing to skyrmion photo-excitation, an explicit description of the electron dynamics is necessary. This will become clear as we critically assess the microscopic mechanism behind skyrmion excitation, and its parameter dependence. In particular, a microscopic analysis of the skyrmion photo-excitation process reveals the $s-d$ exchange and $s$-electron spin-orbit interaction as the key parameters determining the skyrmion excitation probability. This allows for optimization of the skyrmion excitation probability through material engineering, by manipulating these parameters by material choices, optical drives and layer twisting.


{\it Energetics of skyrmion formation.-} 
Starting with the equilibrium properties of the spin system, Fig.~\ref{fig:phase_diagram}$(a)$ shows the magnetic state of the spin system as a function of external magnetic field $B$ and DM interaction $D$. The equilibrium state is found by simulated annealing to a target temperature $k_BT/J = 0.02$ using the Monte Carlo Metropolis algorithm~\cite{Bostrom19}, corresponding to $T = 11.5$ K for $J = 50$ meV. We find three competing equilibrium phases: a spin spiral phase, a FM phase, and a skyrmion crystal (SkX) phase. To excite skyrmions from an initially FM state, we choose the spin parameters $D/J = 0.5$ and $B/J = 0.2$ close to the FM-SkX phase boundary, as indicated by the orange dot in Fig.~\ref{fig:phase_diagram}$(a)$.

To understand how itinerant electrons influence the magnetic state, we calculate the excitation energy $E = E_{Sk} - E_{FM}$ of a single skyrmion on top of the FM state, for a skyrmion radius $R$ and spin-electron coupling $g$ [Fig.~\ref{fig:phase_diagram}$(b)$]. Here $E_{Sk}$ is the energy of the coupled system for the spin configuration ${\bf n}({\bf r}) = (-x f(u), -y f(u), 1-2e^{-u^2})$, where $f(u) = (2u/\sigma)(e^{-u^2}-e^{-2u^2})^{1/2}$, $u = r/R$ and the coordinates are measured from the center of the system. We have verified that that this skyrmion profile agrees well with numerical results for a relaxed skyrmion. As shown in Fig.~\ref{fig:phase_diagram}$(b)$, there is for $g = 0$ an energy barrier of $E \approx 0.75$ eV to excite a skyrmion of finite radius, while for $g = 2.5$ the barrier is reduced to $E \approx 0.15$ eV. In addition, for $g \gtrsim 2$, the skyrmion state is lower in energy than the FM state. Thus, due to the lowering of the energy barrier with increasing $g$, a lower laser fluence is required to excite skyrmions at larger coupling.


{\it The role of spin-orbit interactions.--}
Since a strong $s-d$ exchange is expected to facilitate skyrmion photo-excitation, we can gain further insight into the microscopic mechanism behind the excitation process by deriving an effective equation of motion for the spins in this limit. For large $g$ electrons with spin antiparallel to the localized moments will be penalized by an energy $\sim 2g$, and for $g \to \infty$ the antiparallel component vanishes~\cite{Volovik87}. 

To exploit this fact we adopt a continuum description and perform a local gauge transformation to align the electron spins with the localized moments (for details see SM). Eliminating the antiparallel spin component to lowest order gives a modified Landau-Lifshitz equation
\begin{align}\label{eq:effective_landau}
 \partial_t {\bf n} = - \frac{a^2}{\hbar M} \left( {\bf n} \times [{\bf B}_s + \frac{1}{2} \boldsymbol\alpha \times {\bf j}_e] + \hbar ({\bf j}_s \cdot \nabla) {\bf n} \right).
\end{align}
The prefactor describes the total magnetization density $M = S + ms$, where $m$ is the local spin density and $s = 1/2$, since for $g \to \infty$ the electronic and localized spins form an effective magnetic moment of magnitude $M$. The effective magnetic field ${\bf B}_s$ contains the spin-spin interactions that are renormalized to $J \to J + \rho t/2$ and $D \to D + \rho\alpha/2$ by the interaction with the electrons, where $\rho$ is the density of electrons with spins parallel to the magnetic moments. The term proportional to the charge current ${\bf j}_e$ describes a spin-orbit induced effective magnetic field, which for a Rashba interaction lies in the substrate plane, and the last term describes a coupling to the spin current ${\bf j}_s$.

For a ferromagnetic system ${\bf n}({\bf r}) = {\bf n}$, $\partial_i {\bf n} = 0$ and the last term of Eq.~\ref{eq:effective_landau} vanishes. The dominant coupling between the electrons and magnetic moments is then given by the term proportional to ${\bf j}_e$. For a current ${\bf j}_e = j_e \hat{\bf e}_x$ (as induced by a linearly polarized laser) this term generates a magnetic field ${\bf B}_{so}({\bf r},t) = \alpha j_e \hat{\bf e}_y$ tilting the spins away from the $z$-axis. We note that for a vanishing spin-orbit coupling, the only effect of the electrons is to renormalize the exchange parameter. Thus, without spin-orbit coupling, skyrmion excitation will be strongly suppressed. This has been confirmed in all our simulations, where we always found no skyrmion excitation for $\alpha = 0$ within the parameter range considered.

\begin{figure}
 \includegraphics[width=\columnwidth]{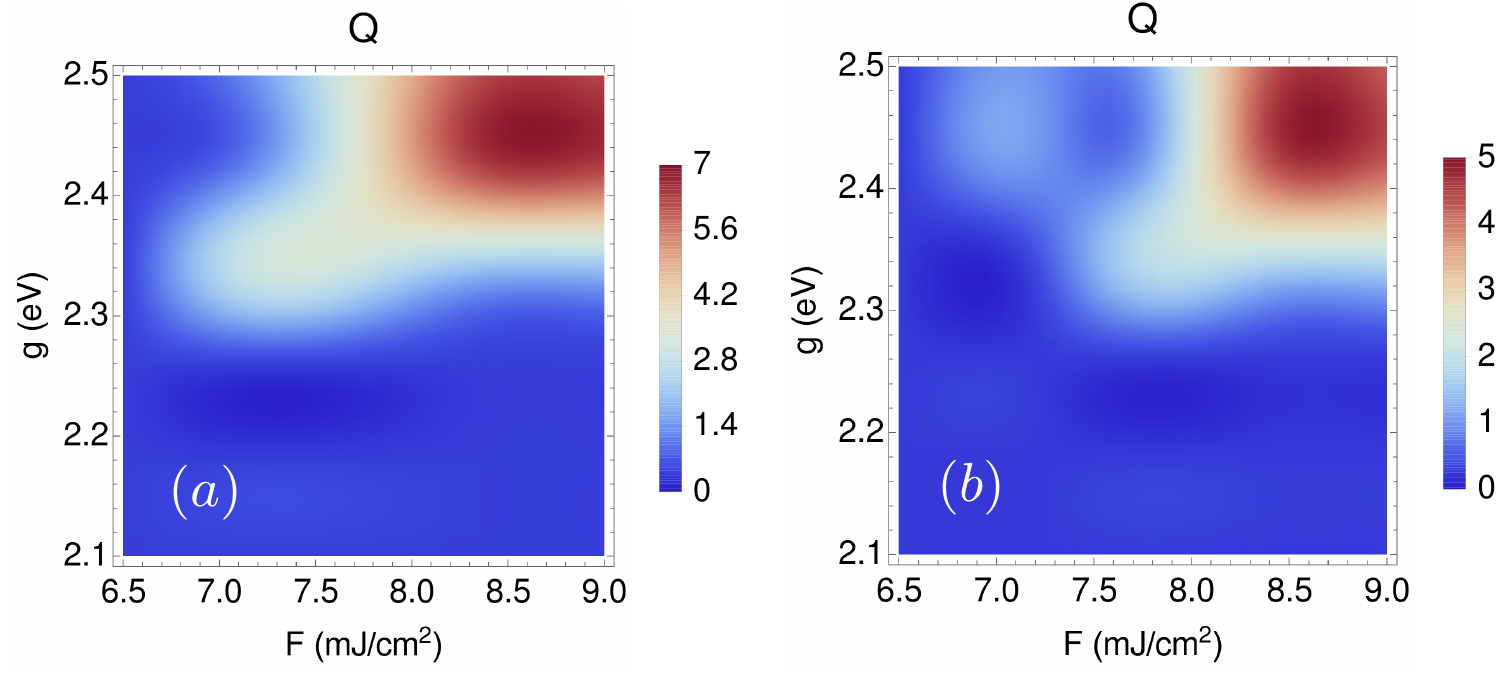}
 \caption{{\bf Parameter dependence of the skyrmion excitation}. Topological charge $Q$ at $t \approx 320$ fs as a function of laser fluence $F$ and spin-electron coupling $g$. The system is square lattice with $30\times 30$ sites with spin parameters $J = 50$ meV and $D = 25$ meV, and electronic parameters $t = 1$ eV and $\alpha = 0.5$ eV. The laser has a photon energy $\hbar\omega = 0.5$ eV, pulse duration $\tau = 30$ fs, and spot size $\approx 12$ nm. Panel $(a)$ shows the topological charge for a magnetic field $B = 7.5$ meV and panel $(b)$ for a $B = 10$ meV.}
 \label{fig:creation_density}
\end{figure}

{\it Parameter dependence of skyrmion excitation and  material realizations.--}
In Fig.~\ref{fig:creation_density} we show the topological charge $Q$ as a function of $F$ and $g$. For values of $g$ smaller than $\approx 2.3$ no skyrmions are excited, while for larger values of $g$ we find the topological charge to be an increasing function of $F$, with an approximately linear dependence. This is in agreement with the expectation from Fig~\ref{fig:phase_diagram}$(b)$: a finite amount of energy, which decreases with $g$, must be supplied in order to overcome the skyrmion excitation barrier. The results also qualitatively agree with the trend found in experiments~\cite{Je20}.

As shown above, favorable conditions to photo-induce skyrmions are a non-zero spin-orbit interaction and large $s-d$ exchange. Since the equilibrium state of the system can be controlled by external magnetic and electric fields~\footnote{The Rashba-induced DM interactions at an interface can be manipulated via a perpendicular electric field}, we assume the equilibrium system is close to the FM-SkX phase boundary. These conditions are likely to be satisfied in transition metal monolayers and thin films~\cite{Fert17}, where the $s-d$ is naturally strong and spin-orbit interactions are enhanced by interfacial inversion symmetry breaking. This includes in particular the well-studied systems Fe/Ir(111) and Pd/Fe/Ir(111)~\cite{Heinze11,Romming15,vonBergmann15}. Another interesting class of systems are twisted magnetic van der Waals bilayers, which have strong spin-orbit interactions and whose local interactions can be tuned via the twist angle~\cite{Tong16}.


{\it Conclusions.--}
We have proposed a microscopic mechanism of laser-induced skyrmion excitation based on a two-band electronic model. The numerical simulations predict ultrafast skyrmion excitation in the limit of strong $s-d$ exchange, which we attribute to the effective magnetic field generated by spin-orbit interactions among the itinerant electrons. The explicit treatment of itinerant electrons and their dynamic interaction with the localized spins is an essential ingredient in this mechanism.

The proposed theory predicts photo-excitation of skyrmions via itinerant electrons on a $100$ fs second timescale, and thus opens up for ultrafast manipulation of topological magnetic textures. In addition, both the $s-d$ exchange and interfacial Rashba interaction are in principle controllable via material choices and external fields, leading to large possibilities of engineering candidate materials such as transition metal thin films and van der Waals bilayer to optimize the skyrmion excitation. Our theory can also be applied to study the influence of itinerant electrons on skyrmion transport~\cite{Bostrom19}, magnon and electron excitations in SkXs~\cite{Daz20,Hamamoto15,Lado15}, and charged skyrmions on topological insulator surfaces~\cite{Nomura10}, thus opening an avenue to explore a vast range of physical phenomena in non-equilibrium magnetic systems.


\begin{acknowledgments}
 {\it Acknowledgments.--} EVB acknowledges stimulating discussions with Florian Eich during the early stages of this work. We acknowledge support by the Max Planck Institute New York City Center for Non-Equilibrium Quantum Phenomena and by the Swedish Research Council. We also acknowledge support by the European Research Council (ERC-2015-AdG694097), the Cluster of Excellence “Advanced Imaging of Matter” (AIM), and Grupos Consolidados (IT1249-19). The Flatiron Institute is a Division of the Simons Foundation.
\end{acknowledgments}

\bibliography{references}

\newpage
\foreach \x in {1,2,3,4,5}
{%
\clearpage


\section*{Supplementary material}

# Microscopic Theory of Ultrafast Skyrmion Excitation by Light: Supplemental Material

Emil Viñas Boström,[1] Angel Rubio,[1, 2] and Claudio Verdozzi[3]

[1] *Max Planck Institute for the Structure and Dynamics of Matter, Luruper Chaussee 149, 22761 Hamburg, Germany*
[2] *Center for Computational Quantum Physics (CCQ), The Flatiron Institute, 162 Fifth Avenue, New York, NY 10010, United States of America*
[3] *Division of Mathematical Physics and ETSF, Lund University, PO Box 118, 221 00 Lund, Sweden*

(Dated: October 29, 2020)

## A. Derivation of the effective spin Hamiltonian

We here provide additional details on the derivation of the effective spin Hamiltonian describing the correlated $d$-electrons in the large $U$ limit. Assuming the $d$-electron system is at half-filling, doubly occupied sites will be penalized by an energy $\sim U$, and the effective Hilbert space can be defined by projecting out the doubly occupied sites. This is achieved by the projection operator $\mathcal{P} = \prod_i \mathcal{P}_i$, where $\mathcal{P}_i = 1 - \hat{n}_{i\uparrow}\hat{n}_{i\downarrow}$. In the following we decompose the Hamiltonian as $H(t) = H_0(t) + H_1$, where $H_1$ is the interacting part of $H_d$ defined in the main text. For virtual excitations out of the subspace defined by $\mathcal{P}$, where exactly one doubly occupied site is involved, we can write (up to an irrelevant constant[1])

$$H_1 = U_0 \sum_i \hat{n}_{i\uparrow}\hat{n}_{i\downarrow} + \frac{V_0}{2}\sum_{\langle ij \rangle} \hat{n}_i \hat{n}_j = U \sum_i \hat{n}_{i\uparrow}\hat{n}_{i\downarrow}, \quad (1)$$

where $U = U_0 - V_0$. We note that $H_1$ only acts in the high energy subspace defined by $\mathcal{Q} = 1 - \mathcal{P}$.

The time-dependent Schrieffer-Wolff transformation is defined as the unitary transformation that at each time $t$ removes the coupling between the low and high energy subspaces[2]. Given a state $|\Psi(t)\rangle$ that evolves under the original Hamiltonian $H(t)$, the unitary transformation $|\tilde{\Psi}(t)\rangle = e^{iS(t)}|\Psi(t)\rangle$ corresponds to a Hamiltonian $\tilde{H} = e^{iS(t)}[H - i\partial_t]e^{-iS(t)}$. Assuming that $S(t)$ can be written as $S(t) = \gamma S_1(t) + \gamma^2 S_2(t) + \mathcal{O}(\gamma^3)$, with $\gamma$ a small parameter, we find to second order in $\gamma$ that

$$\tilde{H} = H_0 + H_1 + \gamma\left(i[S_1, H_0] + i[S_1, H_1] - \partial_t S_1\right) \quad (2)$$
$$+ \gamma^2\left([S_2, H_1] - \frac{1}{2}[S_1, i\partial_t S_1 + [S_1, H_1]] - \partial_t S_2\right).$$

To eliminate the leading order off-diagonal term (in $\mathcal{P}$ and $\mathcal{Q}$), we take $S_2 = 0$ and require that $i\gamma[S_1, H_1] - \gamma\partial_t S_1 = -\mathcal{P}H_0\mathcal{Q} - \mathcal{Q}H_0\mathcal{P}$. Since the projection operators $\mathcal{P}$ and $\mathcal{Q}$ act in the subspace of $d$-electrons, only the $d$-electron kinetic term $T_d$ and the coupling Hamiltonian $H_{s-d}$ contribute to $S_1$. Projecting the expression for $\tilde{H}$ onto the low energy subspace, we have

$$\tilde{H} = \mathcal{P}H_0\mathcal{P} + \frac{i\gamma}{2}\mathcal{P}S_1\mathcal{Q}\mathcal{Q}H_0\mathcal{P} - \frac{i\gamma}{2}\mathcal{P}H_0\mathcal{Q}\mathcal{Q}S_1\mathcal{P}. \quad (3)$$

It now remains to solve the differential equation for $S_1$. This can be achieved by introducing the retarded and advanced Green's functions

$$G^R(t, t') = -ie^{-i(H_1 - i\eta)(t-t')}\theta(t - t') \quad (4a)$$
$$G^A(t, t') = ie^{i(H_1 - i\eta)(t'-t)}\theta(t' - t), \quad (4b)$$

in terms of which the projections of the operator $S_1$ are given by

$$\mathcal{Q}S_1(t)\mathcal{P} = i\int dt'\, G^R(t, t')\mathcal{Q}H_0(t')\mathcal{P} \quad (5a)$$
$$\mathcal{P}S_1(t)\mathcal{Q} = -i\int dt'\, \mathcal{P}H_0(t')\mathcal{Q}G^A(t', t). \quad (5b)$$

Since we work at half-filling the operator $H_1$ appearing in the exponential of the Green's function always acts after a single excitation has been created, and can therefore by replaced by $U$ in the exponential. The effective Hamiltonian is then

$$\tilde{H} \approx \frac{i}{2}\int dt'\, \theta(t - t')\Big[e^{i(U+i\eta)(t-t')}\mathcal{P}H_0(t')\mathcal{Q}\mathcal{Q}H_0(t)\mathcal{P}$$
$$- e^{-i(U-i\eta)(t-t')}\mathcal{P}H_0(t)\mathcal{Q}\mathcal{Q}H_0(t')\mathcal{P}\Big] + H_0. \quad (6)$$

The operator products can be evaluated following the procedure detailed in Section B below, and leads to a time-dependent spin Hamiltonian

$$\tilde{H}(t) = H_0 + \sum_{\langle ij \rangle}\Big[J_{ij}(t)\hat{\mathbf{S}}_i \cdot \hat{\mathbf{S}}_j + \mathbf{D}_{ij}(t) \cdot (\hat{\mathbf{S}}_i \times \hat{\mathbf{S}}_j)$$
$$+ \hat{\mathbf{S}}_{i\mu}\Gamma_{i\mu,j\nu}(t)\hat{\mathbf{S}}_{j\nu}\Big] + H_{s-d}, \quad (7)$$

with the parameters

$$J_{ij}(t) = 4t_d^2 I_{ij}(t) - J_x \quad (8a)$$
$$\mathbf{D}_{ij}(t) = 8it_d\boldsymbol{\alpha}_d I_{ij}(t) \quad (8b)$$
$$\Gamma_{i\mu,j\nu}(t) = (8\alpha_{d,\mu}\alpha_{d,\nu} + 4|\boldsymbol{\alpha}_d|^2\delta_{\mu\nu})I_{ij}(t). \quad (8c)$$

Here we use Greek letters to denote the components of the spin-orbit vector $\boldsymbol{\alpha}$. The time-dependent function $I_{ij}(t)$ is given by

$$I(t) = \text{Im}\int dt'\, e^{i(U+i\eta)(t-t')}\cos(\theta_{ij}(t) - \theta_{ij}(t')). \quad (9)$$

Similarly the $s-d$ exchange interaction (obtained from the time-independent part of Eq. 6) is found to be of the form

$$H_{s-d} = -g \sum_i \hat{\mathbf{s}}_i \cdot \hat{\mathbf{S}}_i, \qquad (10)$$

where $g = J_{s-d} - 4t_{s-d}^2/U$.

### B. Evaluation of the operator products of the effective Hamiltonian

Given the formal expression for $\tilde{H}$ in Eq. 6, we can evaluate the operator products following the procedure in Ref.[3]. At half-filling only virtual transitions that involve two sites contribute to the Hamiltonian, and thus the projection operators $\mathcal{Q}$ always give unity and can be removed. The products arising from the kinetic energy of the $d$-electrons are then of the form

$$T_d(t')T_d(t) \qquad (11)$$
$$= \sum_{\langle ij \rangle} M^{ij}_{\sigma_1\sigma_2}(t')(\delta_{\sigma_2\sigma_3} - d^\dagger_{j\sigma_3}d_{j\sigma_2})M^{ji}_{\sigma_3\sigma_4}(t)d^\dagger_{i\sigma_1}d_{i\sigma_4},$$

where summation over repeated spin indexes is implied and the projection operators $\mathcal{P}$ have been left out for notational simplicity. We now use that the electronic bilinears can be represented in terms of Pauli matrices as $d^\dagger_{i\sigma}d_{i\sigma'} = (1/2 + \hat{\mathbf{S}}_i \cdot \boldsymbol{\tau})_{\sigma'\sigma}$. Similarly, the products arising from the $s-d$ exchange term are of the form

$$H_{s-d}H_{s-d} = t_{s-d}^2 \sum_i (\delta_{\sigma_1\sigma_2} - c^\dagger_{i\sigma_2}c_{i\sigma_1})d^\dagger_{i\sigma_1}d_{i\sigma_2}, \qquad (12)$$

where we can represent the $s$-electron bilinear by $c^\dagger_{i\sigma}c_{i\sigma'} = (1/2 + \hat{\mathbf{s}}_i \cdot \boldsymbol{\tau})_{\sigma'\sigma}$.

To this order in the Schrieffer-Wolff expansion no other terms can arise, and in particular, terms of the form $\mathcal{P}T_dH_{s-d}\mathcal{P} = 0$ since they contain an odd number of $d$ operators. Therefore, they necessarily create (or destroy) a doubly occupied state, which takes the system out of the low-energy Hilbert space.

With the above representation of the electronic operators the expression for $\tilde{H}$ takes the form of a trace, explicitly given by

$$\tilde{H} \approx H_0 + \sum_{\langle ij \rangle} \text{Tr} \left[ \mathbf{M}^{ij}(t')\mathbf{M}^{ji}(t) \left( \frac{1}{2} + \hat{\mathbf{S}}_i \cdot \boldsymbol{\tau} \right) \right] \qquad (13)$$
$$- \sum_{\langle ij \rangle} \text{Tr} \left[ \mathbf{M}^{ij}(t') \left( \frac{1}{2} + \mathbf{S}_j \cdot \boldsymbol{\tau} \right) \mathbf{M}^{ji}(t) \left( \frac{1}{2} + \mathbf{S}_i \cdot \boldsymbol{\tau} \right) \right]$$
$$+ t_{s-d}^2 \sum_i \text{Tr} \left[ \frac{1}{2} + \hat{\mathbf{S}}_i \cdot \boldsymbol{\tau} \right]$$
$$- t_{s-d}^2 \sum_i \text{Tr} \left[ \left( \frac{1}{2} + \hat{\mathbf{s}}_i \cdot \boldsymbol{\tau} \right) \left( \frac{1}{2} + \hat{\mathbf{S}}_i \cdot \boldsymbol{\tau} \right) \right]$$

We evaluate these traces using a representation of $\mathbf{M}$ in terms of the Pauli matrices, $\mathbf{M} = -t_d\mathbf{1} + \boldsymbol{\alpha}_d \cdot \boldsymbol{\tau}$, together with the property $(\mathbf{a} \cdot \boldsymbol{\tau})(\mathbf{b} \cdot \boldsymbol{\tau}) = (\mathbf{a} \cdot \mathbf{b})\mathbf{1} + i(\mathbf{a} \times \mathbf{b}) \cdot \boldsymbol{\tau}$ and the trace identities $\text{Tr}\,\mathbf{1} = 2$ and $\text{Tr}\,\boldsymbol{\tau} = 0$. The first term in $\tilde{H}$ (for given $i$ and $j$) can be shown to be independent of the spin vectors for real hoppings $t_d^* = t_d$ and imaginary spin-orbit couplings $\boldsymbol{\alpha}_d^* = -\boldsymbol{\alpha}_d$, which holds when the spin-orbit interaction is of the Rashba form. This term can therefore be neglected. The same argument holds for the terms proportional to the identity matrix in the second, third and fourth term, which can thus be omitted too. The remaining part of the trace is then

$$\text{Tr}\left[\mathbf{M}^{ij}(t')(\mathbf{S}_j \cdot \boldsymbol{\tau})\mathbf{M}^{ji}(t)(\mathbf{S}_i \cdot \boldsymbol{\tau})\right] = e^{i(\theta_{ij}(t') - \theta_{ij}(t))}$$
$$\times \left[2t_d^2 \mathbf{S}_i \cdot \mathbf{S}_j + 2i(t_d \boldsymbol{\alpha}_d^* - t_d \boldsymbol{\alpha}_d) \cdot (\mathbf{S}_i \times \mathbf{S}_j) \right.$$
$$\left. - 2(\boldsymbol{\alpha}_d \times \mathbf{S}_i) \cdot (\boldsymbol{\alpha}_d^* \times \mathbf{S}_j) + 2(\boldsymbol{\alpha}_d \cdot \mathbf{S}_i)(\boldsymbol{\alpha}_d^* \cdot \mathbf{S}_j)\right].$$

for the kinetic part, and

$$\text{Tr}\left[t_{s-d}^2(\mathbf{s}_i \cdot \boldsymbol{\tau})(\mathbf{S}_i \cdot \boldsymbol{\tau})\right] = 2t_{s-d}^2 \mathbf{s}_i \cdot \mathbf{S}_i. \qquad (14)$$

for the $s-d$ exchange part. Inserting these results in the expression for $\tilde{H}$, and noting that the term with $i$ and $j$ swapped gives an analogous contribution to the kinetic part but with the phase $e^{-i(\theta_{ij}(t') - \theta_{ij}(t))}$, we find the effective spin Hamiltonian of Eq. (7). The final expression for $I_{ij}(t)$ comes from combining the terms in Eq. (6) relating to the retarded and advanced Green's functions. Similarly, the extra factor of two in Eq. 10 for the $s-d$ exchange comes from the Hermitian conjugate of the term considered above.

### C. Definition of the topological charge

To quantify the topology of the classical spin texture we use the topological charge $Q$, defined on a lattice by[4]

$$Q = \frac{1}{4\pi} \sum_\Delta \Omega_\Delta. \qquad (15)$$

In this definition the lattice is triangulated and $\Omega_\Delta$ corresponds to the signed area of the spherical triangle spanned by three neighboring spins, given by[4]

$$e^{i\Omega_\Delta/2} = \frac{1}{\rho}(1 + \mathbf{S}_i \cdot \mathbf{S}_j + \mathbf{S}_j \cdot \mathbf{S}_k + \mathbf{S}_k \cdot \mathbf{S}_i \qquad (16)$$
$$+ i\eta_{ijk}\mathbf{S}_i \cdot [\mathbf{S}_j \times \mathbf{S}_k])$$
$$\rho = \sqrt{2(1 + \mathbf{S}_i \cdot \mathbf{S}_j)(1 + \mathbf{S}_j \cdot \mathbf{S}_k)(1 + \mathbf{S}_k \cdot \mathbf{S}_i)},$$

where $\eta_{ijk} = +1\,(-1)$ if the path $i \to j \to k \to i$ is positively (negatively) oriented. The surface area $\Omega_\Delta$ is well-defined everywhere except at the zero-measure set $\mathbf{S}_i \cdot (\mathbf{S}_j \times \mathbf{S}_k) = 0$ and $1 + \mathbf{S}_i \cdot \mathbf{S}_j + \mathbf{S}_j \cdot \mathbf{S}_k + \mathbf{S}_k \cdot \mathbf{S}_i < 0$, where $e^{i\Omega_\Delta/2}$ has a branch cut. The topological charge is a compact, convenient indicator of the presence of a nontrivial spin texture: For a single skyrmion $Q = -1$, for a single antiskyrmion $Q = 1$, and for a cluster of skyrmions and antiskyrmions $Q = \sum_i Q_i$.

### D. Inefficiency of the inverse Faraday effect for atomically thin samples

A mechanism proposed to underlie the optical excitation of skyrmions is the direct coupling between the material magnetization and the laser electric field via the inverse Faraday effect (IFE)[5,6]. However, a straightforward estimate of the interaction energies involved in the IFE shows that it can not be responsible for skyrmion excitation in quasi two-dimensional systems. The interaction energy for the IFE in a volume $a^3$ can be written as[7]

$$U = -\frac{i\theta_F c \sqrt{\epsilon_r} \epsilon_0 a^3}{2\omega} \frac{\mathbf{M}(\mathbf{r})}{M_s} \cdot [\mathbf{E}^*(\mathbf{r}) \times \mathbf{E}(\mathbf{r})], \quad (17)$$

where $\theta_F$ is the Faraday angle, $\epsilon_r$ the relative permittivity of the material, $a$ the lattice parameter, $\omega$ the frequency of the laser and $M_s$ the saturation magnetization. The Faraday angle can be written as $\theta_F = \mathcal{V}B$ where $\mathcal{V}$ is the so-called Verdet constant, which is smaller than $\sim 100$ rad/Tm. Taking a large value $E = 10^9$ V/m, $a = 5$ Å and $\lambda = 800$ nm, giving $\omega \approx 2360$ THz, we find the energy density $U = g\mathbf{S} \cdot \mathbf{e}$ with $g \approx 1.2 \cdot 10^{-7}$ eV, $\mathbf{S} = \mathbf{M}/M_s$ and $\mathbf{e} = (\mathbf{E}^* \times \mathbf{E})/E^2$. Since typical values of spin parameters are on the order of $\sim 1$ meV, the IFE coupling is at least a factor $10^{-3}$ smaller than the direct Zeeman term. The reason for this small coupling is that the strength of the IFE is proportional to the propagation length through the system[7], and is therefore strongly suppressed for atomically thin systems. In these systems, the IFE is to a very good approximation negligible.

### E. Spin equation of motion in the large $s-d$ exchange limit

To better understand the influence of itinerant electrons on the dynamics of the spin system, we derive an effective equation of motion for the spins in the limit of large $s-d$ exchange[8–12]. To simplify the algebraic manipulations we adopt a continuum description $\psi_\sigma(\mathbf{r}_i) = c_{i\sigma}/a$ and $\mathbf{n}(\mathbf{r}_i) = \mathbf{n}_i$, where the Lagrangian for the itinerant electrons is given by

$$\mathcal{L} = \int d^2\mathbf{r} \left( \psi_\sigma^\dagger \left[ i\hbar\partial_t + \frac{\hbar^2}{2m}\nabla^2 - \frac{i}{2}(\boldsymbol{\alpha}_i \cdot \boldsymbol{\tau}) \overset{\leftrightarrow}{\nabla}_i \right] \psi_\sigma + g\psi_\sigma^\dagger (\boldsymbol{\tau} \cdot \mathbf{n}) \psi_{\sigma'} \right). \quad (18)$$

Here $\psi_\sigma^\dagger \overset{\leftrightarrow}{\nabla} \psi_\sigma = \psi_\sigma^\dagger (\nabla \psi_\sigma) - (\nabla \psi_\sigma^\dagger)\psi_\sigma$, and the spin-orbit interaction has been written in a general form in order to facilitate the algebraic manipulations. We note that here $\alpha_i^a$ is a matrix with $i$ denoting the spatial and $a$ the spin component, which reduces to the standard Rashba spin-orbit coupling at an interface, $H_{so} = \alpha \hat{\mathbf{e}}_z \cdot (\nabla \times \boldsymbol{\tau})$, by taking $\boldsymbol{\alpha}_x = \alpha \hat{\mathbf{e}}_y$, $\boldsymbol{\alpha}_y = -\alpha \hat{\mathbf{e}}_x$ and $\boldsymbol{\alpha}_z = 0$.

We assume the magnetization is described by a normalized spin texture $\mathbf{n}(\mathbf{r}, t) = (\sin\theta\cos\phi, \sin\theta\sin\phi, \cos\theta)$ where the angles $\theta = \theta(\mathbf{r}, t)$ and $\phi = \phi(\mathbf{r}, t)$ are functions of space and time. Exploiting the gauge invariance of the theory, we perform a local $SU(2)$ transformation $\psi_\sigma(\mathbf{r}, t) \to U_{\sigma\sigma'}(\mathbf{r}, t)\psi_{\sigma'}(\mathbf{r}, t)$ of the electronic field operators to align the electronic spins with the underlying magnetic texture. The transformation corresponds to a local rotation and is implemented by the operator $U = \mathbf{m} \cdot \boldsymbol{\tau}$ for $SU(2)$ vectors and by the matrix $R_{ab} = 2m_a m_b - \delta_{ab}$ for $SO(3)$ vectors, where $\mathbf{m}(\mathbf{r}, t) = (\sin\frac{\theta}{2}\cos\phi, \sin\frac{\theta}{2}\sin\phi, \cos\frac{\theta}{2})$.

Under the above gauge rotation the derivatives transform like $\partial_\mu \psi_\sigma = U(\partial_\mu + iA_\mu)\psi_\sigma$, where $A_\mu = -iU^\dagger \partial_\mu U$ acts like an emergent electromagnetic field. The Lagrangian for the itinerant electrons then becomes

$$\mathcal{L} = \int d^2\mathbf{r} \left( i\hbar \psi_\sigma^\dagger (\partial_t + iA_0)\psi_\sigma + g\psi_\sigma^\dagger \tau^z \psi_{\sigma'} \right.$$
$$+ \sum_i \left[ \frac{\hbar^2}{2m} \psi_\sigma^\dagger (\nabla_i^2 + iA_i \overset{\leftrightarrow}{\nabla}_i - A_i^2)\psi_{\sigma'} \right. \quad (19)$$
$$\left. \left. - \frac{i}{2} \psi_\sigma^\dagger (\boldsymbol{\alpha}_i' \cdot \boldsymbol{\tau}) \overset{\leftrightarrow}{\nabla}_i \psi_{\sigma'} + \frac{1}{2} \psi_\sigma^\dagger (\boldsymbol{\alpha}_i' \cdot \boldsymbol{\tau}) A_i \psi_\sigma \right] \right),$$

and we note that the exchange interaction is now diagonal. Instead the coupling between the spins and electrons is mediated via the gauge fields and the rotated spin-orbit interaction $\alpha_i^{'a} = \sum_a R_{ab}\alpha_i^b$.

We can decompose the Lagrangian into a part independent of the local magnetization and a part which depends on $A_\mu$ and $\boldsymbol{\alpha}_i$. The first part is

$$\mathcal{L}_0 = \int d^2\mathbf{r} \left[ \psi_\sigma^\dagger \left( i\hbar\partial_t + \frac{\hbar^2}{2m}\nabla^2 \right) \psi_\sigma + g\psi_\sigma^\dagger \tau^z \psi_{\sigma'} \right], \quad (20)$$

which describes free electrons in presence of a magnetic field $\mathbf{B} = g\hat{\mathbf{e}}_z$. The second part of the Lagrangian is

$$\mathcal{L}_A = -\int d^2\mathbf{r} \left[ \hbar a_0 \rho_\sigma + \frac{i}{2} \sum_i \left( \psi_\sigma^\dagger (\boldsymbol{\alpha}_i' \cdot \boldsymbol{\tau}) \overset{\leftrightarrow}{\nabla}_i \psi_{\sigma'} \right. \right. \quad (21)$$
$$\left. \left. - \frac{i\hbar^2}{2m} \psi_\sigma^\dagger (\mathbf{a}_i \cdot \boldsymbol{\tau}) \overset{\leftrightarrow}{\nabla}_i \psi_{\sigma'} - \frac{1}{2}(\boldsymbol{\alpha}_i' \cdot \mathbf{a}_i)\rho_\sigma + \frac{\hbar^2}{2m}\mathbf{a}_i^2 \rho_\sigma \right) \right],$$

where $\rho_\sigma = \psi_\sigma^\dagger \psi_\sigma$ is the electronic spin density operator and we have written $A_0 = a_0 \mathbf{1}$ and $A_i = \mathbf{a}_i \cdot \boldsymbol{\tau}$.

We note that the last three terms of $\mathcal{L}_A$ can be combined into a coupling between the electronic spin current $\mathbf{j}_i$ and the emergent electromagnetic field $\mathbf{a}_i$, written on the form

$$\mathcal{L}_c = -\int d^2\mathbf{r} \sum_i \mathbf{j}_i \cdot \mathbf{a}_i \quad (22)$$

$$\mathbf{j}_i = -\frac{i\hbar^2}{2m} \psi_\sigma^\dagger \boldsymbol{\tau} \overset{\leftrightarrow}{\nabla}_i \psi_{\sigma'} + \left( \frac{\hbar^2}{2m}\mathbf{a}_i - \frac{1}{2}\boldsymbol{\alpha}_i' \right)\rho_\sigma \quad (23)$$

Due to the gauge transformation the spin components of the current are given in the local frame specified by the operator $U$. Rotating $\mathcal{L}_c$ back to the laboratory frame and using the fact that $R_{ab}a_i^b = (\nabla_i \mathbf{n} \times \mathbf{n})^a + n^a a_i^{z\,[10]}$, we find the coupling

$$\mathcal{L}_c = -\int d^2\mathbf{r} \sum_i \left( \mathbf{D}_i \cdot (\nabla_i \mathbf{n} \times \mathbf{n}) + j_i^\parallel a_i^z \right). \quad (24)$$

The first term describes a Dzyaloshinskii-Moriya (DM) coupling between the localized spins, with a DM vector given by $\mathbf{D}_i = \mathbf{j}_{\perp,i} = \mathbf{j}_i - j_{\parallel,i}$. Here $\mathbf{j}_i$ is the current in the laboratory frame and $j_{\parallel,i} = \mathbf{n} \cdot \mathbf{j}_i$ is the spin component of the current parallel to the magnetic texture. The parallel component can be subtracted from $\mathbf{D}_i$ since $\mathbf{n} \cdot (\nabla_i \mathbf{n} \times \mathbf{n}) = 0$.

So far the algebraic manipulations have been exact and the Lagrangian $\mathcal{L} = \mathcal{L}_0 + \mathcal{L}_A$ gives an exact reformulation of the initial problem. To derive an effective Lagrangian for large values of $g$ we note that for $g \to \infty$ the electronic spin component antiparallel to the local magnetization will be strongly suppressed. In this limit we can write the spin-orbit interaction like $\boldsymbol{\alpha}' \cdot \boldsymbol{\tau} \approx (\boldsymbol{\alpha} \times \mathbf{n})\tau^z$. Noting that $\mathbf{a}_i^2 = \frac{1}{4}(\partial_i \mathbf{n})^2 + (a_i^z)^2$[11], and writing $a_i^z = \mathbf{a}$, the effective spin Lagrangian can be written

$$\mathcal{L}_s = -\int d^2\mathbf{r} \left[ \frac{2S\hbar}{a^2} a_0 + H_s + \hbar a_0 \rho \right.$$
$$+ \frac{\hbar^2 \rho}{8m}(\partial_i \mathbf{n})^2 + \sum_i \mathbf{D}_i \cdot (\nabla_i \mathbf{n} \times \mathbf{n}) \quad (25)$$
$$\left. + \mathbf{j}_\parallel \cdot \mathbf{a} + \frac{i}{2}(\boldsymbol{\alpha} \times \mathbf{n}) \cdot \psi^\dagger \tau_z \overset{\leftrightarrow}{\nabla} \psi \right]$$

Here the first two terms give the Lagrangian of the isolated spin system, and $\psi$ and $\rho$ are the field operator and density operator for the spin component parallel to the local magnetization. We note that in the strong coupling limit the emergent vector potential is given by $a_\mu = (a_0, \mathbf{a}) = \frac{1}{2}(1 - \cos\theta)\partial_\mu \phi$.

The equations of motion for the spin system are obtained by replacing the electronic operators by their averages and varying the Lagrangian. Using the relation $\delta a_\mu / \delta \mathbf{n} = \frac{1}{2}(\partial_\mu \mathbf{n}) \times \mathbf{n}$[11] we find

$$\left( \frac{\hbar S}{a^2} + \frac{\hbar \rho}{2} \right)(\partial_t \mathbf{n}) \times \mathbf{n} + \frac{\delta \mathcal{H}}{\delta \mathbf{n}} + ([\mathbf{j}_s \cdot \nabla]\mathbf{n}) \times \mathbf{n} = 0, \quad (26)$$

where the contribution from the effective spin Hamiltonian is

$$\frac{\delta \mathcal{H}}{\delta \mathbf{n}} = \frac{\delta}{\delta \mathbf{n}} \left( H_s + \frac{\hbar^2 \rho}{8m}(\partial_\mu \mathbf{n})^2 + \mathbf{D}^i \cdot (\partial_i \mathbf{n} \times \mathbf{n}) \right)$$
$$+ \frac{1}{2} \boldsymbol{\alpha} \times \mathbf{j}_e. \quad (27)$$

Here $\mathbf{j}_e = \psi^\dagger \overset{\leftrightarrow}{\nabla} \psi$ and the second and third terms in the first line renormalize the exchange and DM interactions to $J \to J + \hbar^2 \rho/(4m)$ and $\mathbf{D}_i \to \mathbf{D}_i + \mathbf{j}_{\perp,i}$. The current $\mathbf{j}_s$ is given by

$$\mathbf{j}_s = -\frac{i\hbar^2}{2m} \psi^\dagger \overset{\leftrightarrow}{\nabla} \psi + \left( \frac{\hbar^2}{m}\mathbf{a} + \frac{1}{2}\boldsymbol{\alpha} \times \mathbf{n} \right)\rho. \quad (28)$$

Multiplying from the right by $\times \mathbf{n}$ we arrive at the modified Landau-Lifshitz equation

$$\partial_t \mathbf{n} = -\left( \frac{\hbar S}{a^2} + \frac{\hbar \rho}{2} \right)^{-1} \left( \mathbf{n} \times \frac{\delta \mathcal{H}}{\delta \mathbf{n}} + \hbar(\mathbf{j}_s \cdot \nabla)\mathbf{n} \right). \quad (29)$$

We see that apart from the renormalization of the exchange and DM interactions discussed above, the electrons affect the spin dynamics in the following ways: (i) The prefactor describes the total magnetization density (instead of the density of local moments), with $\rho = \langle \psi^\dagger \psi \rangle$ the density of electrons with spins parallel to the local moments. This can be understood by observing that in the $g \to \infty$ limit the electronic spins and localized moments get locked in a parallel configuration, forming a magnetic moment of magnitude $M = S/a^2 + \rho/2$. (ii) The spin-orbit coupling contributes an effective magnetic field $\mathbf{B}_{so} = (1/2)\boldsymbol{\alpha} \times \mathbf{j}_e$ to the spin Hamiltonian, which for a Rashba type interaction is parallel to the plane of the spins. (iii) The localized moments couple to the parallel component of the spin current via the last term in Eq. 29.

If we start from a ferromagnetic state $\mathbf{n} = \hat{\mathbf{e}}_z$ the last term of Eq. 29 is zero. Unless the system spontaneously reorders due to the renormalization of the spin parameters (which happens on long time scales), it is therefore necessary with a non-zero spin-orbit coupling in order to tilt the spins away from their ferromagnetic alignment.

We end this section by calculating the equilibrium renormalization of the DM interaction. Assuming that the first two terms of the spin current $\mathbf{j}_i$ vanish in the ground state, we have $\mathbf{D}_i = -(\rho/2)\boldsymbol{\alpha}_i$ (remembering that the spin current is given in the laboratory frame). We thus find $\mathbf{D}_x = -D\hat{\mathbf{e}}_y$, $\mathbf{D}_y = D\hat{\mathbf{e}}_x$ and $\mathbf{D}_z = 0$ with $D = \rho\alpha/2$, and the DM term in the effective Hamiltonian can be written as

$$\mathbf{D}^i \cdot (\partial_i \mathbf{n} \times \mathbf{n}) = D(\mathbf{n} \cdot \boldsymbol{\partial})n_z - Dn_z(\boldsymbol{\partial} \cdot \mathbf{n}), \quad (30)$$

where $\boldsymbol{\partial} = (\partial_x, \partial_y, 0)$. This is the continuum version of a DM Hamiltonian of the Néel type. The renormalized DM interaction in equilibrium is therefore obtained by the replacement $D \to D + \rho\alpha/2$.

---


}

\end{document}